\documentclass[12pt]{article}
\usepackage{epsfig}
\usepackage{a4,isolatin1}
\newcommand{\Ibb}[1]{ {\rm I\ifmmode\mkern 
            -3.6mu\else\kern -.2em\fi#1}}
\newcommand{\ibb}[1]{\leavevmode\hbox{\kern.3em\vrule
     height 1.2ex depth -.3ex width .2pt\kern-.3em\rm#1}}

\newcommand{\Cx}{{\ibb C}}                                          
                                          
\newcommand{\bi}{b_{ik}}
\newcommand{\beq}{\begin{equation}}
\newcommand{\eeq}{\end{equation}}
\newcommand{\di}{d_{ik}}
\newcommand{\dk}{d_{ki}}
\newcommand{\Hi}{{\it H}}
\newcommand{\Hia}{{\it H_0}}
\newcommand{\Hib}{{\it H_1}}

\newcommand{\Hsb}{{\it H'_1}}
\newcommand{\Fc}{\cal F}

\newcommand{\dast}{d^{\ast}}

\begin{document}
\begin{center}
\vspace*{1.0cm}

{\LARGE{\bf A New Approach to Functional Analysis on Graphs, the
    Connes-Spectral Triple and its Distance Function}} 

\vskip 1.5cm

{\large {\bf Manfred Requardt }} 

\vskip 0.5 cm 

Institut f\"ur Theoretische Physik \\ 
Universit\"at G\"ottingen \\ 
Bunsenstrasse 9 \\ 
37073 G\"ottingen \quad Germany

\end{center}

\vspace{1 cm}

\begin{abstract}
We develop a certain piece of functional
analysis on general graphs and use it to create what Connes calls
a'{\it spectral triple}', i.e. a Hilbert space structure, a
representation of a certain (function) algebra and a socalled '{\it
 Dirac operator}', encoding part of the geometric/algebraic
properties of the graph. We derive in particular an explicit
expression for the '{\it Connes-distance function}' and show that it
is in general bounded from above by the ordinary distance on graphs
(being, typically, strictly smaller(!) than the latter). We exhibit,
among other things, the underlying reason for this phenomenon.
\end{abstract} \newpage

\section{Introduction}
In \cite{1} we developed a version of discrete cellular network
dynamics which is designed to mimic or implement certain aspects of
Planck-scale physics (see also \cite{2}). Our enterprise needed the
developement of a variety of relatively advanced and not entirely
common mathematical tools to cope with such complex and irregular
structures as, say, {\it random graphs} and general networks.

On the mathematical side two central themes are the creation of an
appropriate discrete analysis and a kind of discrete (differential)
geometry or topology. The latter one comprises, among other things, e.g. a version of
dimension theory for such objects (see \cite{3}). It is the former one
which turned out to be related to Connes' {\it noncommutative
  geometry} (see e.g. \cite{4}) insofar as differential calculus on
graphs (which are the underlying geometric structure of our physically
motivated networks) is '{\it non-local}', i.e. mildly non-commutative
(for more details see \cite{1}).

Graphs carry a natural {\it metric structure} given by a '{\it
  distance function}' $d(x,y)$, $x,y$ two {\it nodes} of the graph
  (see the following sections) and which we employed in e.g. \cite{3}
  to develop dimensional concepts on graphs. Having Connes' concept of
  distance in noncommutative geometry in mind (cf. chapt. VI of
  \cite{4}), it is a natural question to compute it in model systems,
  which means in our context: arbitrary graphs, and compare it with the already existing
  notion of graph distance mentioned above.

To this end one has, in a first step, to construct what Connes calls a
'{\it spectral triple}', in other words our first aim is it to recast
part of the (functional) analysis on graphs (developed e.g. in
\cite{1} and in particular in section 3 of this paper as far as
operator theory is concerned) in order to get both an interesting and natural Hilbert space structure , a
corresponding representation of a certain (function) algebra and a
natural candidate for a socalled '{\it Dirac operator}' (not to be
confused with the ordinary Dirac operator of the Dirac equation),
which has to encode certain properties of the '{\it graph
  algebra}'. This will be done in section 4.

In the last section, which deals with the distance concept deriving
from the spectral triple, we will give this notion a closer inspection
as far as graphs and similar spaces are concerned. In this connection
some recent work should be mentioned, in which Connes' distance
function was analyzed on certain simple models like
e.g. one-dimensional lattices (\cite{5}-\cite{7}). These papers
already show that it is a touchy business to isolate ''the'' correct
Dirac operator and that it is perhaps worthwhile to scrutinize the
whole topic in a more systematic way.

One of the advantages of our approach is that it both establishes a
veritable piece of interesting functional analysis on graphs and yields a clear recipe
how to calculate the '{\it Connes distance}' in the most general cases
,exhibiting its true nature in this context as a non-trivial
constraint on certain function classes on graphs. From this latter
remark one can already conclude that its relation to the ordinary
distance is far from trivial! We proved in fact that the
Connes-distance is always bounded by the ordinary distance (being
typically strictly smaller) and clarified the underlying reason for this.

As to the mentioned papers \cite{5} to \cite{7} we would like to say
that we realized their existence only after our manuscript has been
completed, which should be apparent from the fact that our approach is
sufficiently distinct. Therefore we prefer to develop our own line of
reasonong in the paper under discussion and compare it in more detail
with the other approaches elsewhere.

\section{A brief Survey of Differential Calculus on\\ Graphs}
The following is a very sketchy compilation of certain concepts needed
in the further analysis and may partly be new or known only to some
experts in the field (even for specialists on graph theory the '{\it
  functional analysis-point of view}' is perhaps not entirely
common). Our personal starting point is laid down in \cite{1} and
\cite{2}. Two beautiful monographs about certain aspects of functional
analysis on graphs (mostly on the level of finite matrix theory,
i.e. typically being applied to finite graphs) are
\cite{8},\cite{9}. As we are presently developing this fascinating
field further into various directions in a more systematic way we
refer the reader to forthcoming work as to more details.\\[0.5cm]
{\bf 2.1 Definition (Simple Graph)}: i) A {\it simple graph} $G$
consists of a countable set of ({\it labelled}) {\it nodes} or {\it
  vertices} ${\cal V}$, elements denoted by e.g. $n_i$ or $x,y\ldots$ and a set
of {\it bonds} or {\it edges} ${\cal E}$ with {\cal E} isomorphic to a
subset of ${\cal V}\times{\cal V}$. We exclude the possibility of {\it
  elementary loops}, i.e. the above relation is {\it
  nonreflexive}. Stated differently, with $x,y$ some nodes, $(x,y)\in
{\cal E}$ implies $x\neq y$. Furthermore two nodes are connected by
atmost one bond.\\
ii) In \cite{1} we preferred to give the bonds an {\it orientation},
i.e. we can orient the bond connecting the nodes $n_i,n_k$ by
expressing it as $\bi$ or $b_{ki}$ (this is not to be confused with
the notion of {\it directed bonds}; see below). Concepts like these
aquire their full relevance when the graph is given an {\it algebraic
  structure} (\cite{1}).\\[0.5cm]
{\bf 2.2 Algebraic Orientation}: In a sense to be specified below we
postulate 
\beq \bi=-b_{ki}\eeq
{\bf 2.3 First Step of a Differential Structure}: i) We can now define
the\\ {\it Node(Vertex)-Space} $C_0$ and the {\it Bond(Edge)-Space}
$C_1$ by considering functions from ${\cal V}$ and ${\cal E}$ to
certain spaces, in the most simplest case the complex numbers
$\Cx$. (Note that many other choices are possible; the above is only
the choice we make for convenience in the following!). As our basic
sets are discrete, the elementary building blocks are the {\it
  indicator functions} $n_i$,$\bi$ itself (see \cite{1}), i.e. which
have the value $1$ on $n_i$ or $\bi$ and being zero elsewhere.\\
{\bf Consequence}: $C_0,C_1$ can be regarded as (complex) vector
spaces with basis elements $n_i,\bi$.\\
ii) We introduce two operators $d,\delta$, well-known from algebraic
topology (see \cite{1} for more motivation). In a first step they are
given on the basis elements  and then extended linearly:
\beq \delta:\; \bi\;\to\;n_k-n_i\eeq
\beq d:\; n_i\; \to\;\sum_k b_{ki}\eeq
the sum running over the nodes $n_k$ being directly connected with
$n_i$ by a bond. We see that $\delta$ maps $C_1$ into $C_0$ while $d$
maps $C_0$ into $C_1$.\\[0.5cm]
{\bf 2.4 Observation}: That $d$ has in fact the character of a
differential operation can be seen by the following identity
(\cite{1}):
\beq df=1/2\cdot\sum_{ik} (f_k-f_i)\bi\eeq
where the factor $1/2$ arises only from the symmetric summation over
$i$ and $k$ which counts (for convenience) each bond twice on the rhs
and where the above relation $\bi=-b_{ki}$ has been
employed.\vspace{0.5cm}

As long as we remain on the level of pure vector spaces the above
choice is probably the most natural one. But if we try to make the
framework into a full differential calculus (e.g. in the spirit of
noncommutative geometry) we need something like a (left-,right-)module
structure, i.e. we have to multiply elements from $C_1$ from the
left/right with functions from $C_0$ in order to arrive at something
what is called a {\it differential algebra} (with e.g. a {\it Leibniz rule}
to hold). These things have been discussed in more detail in \cite{1};
for a different approach see also \cite{10}, which is more in the
spirit of Connes (\cite{4}).

To this end we have to extend or rather embed the space $C_1$ (in)to a
larger $C'_1$ with basis elements denoted by us as $d_{ik}$.\\[0.5cm]
{\bf 2.5 Definition/Observation}: For various reasons (\cite{1}) we
have to enlarge the space $C_1$ to $C'_1$ with new basis elements
$d_{ik}$, but now with $d_{ik}$ being linearly independent from
$d_{ki}$ and
\beq \bi:=d_{ik}-d_{ki}\eeq
a) Pictorially $d_{ik}$, $d_{ki}$ may be considered as {\it directed
  bonds} (in contrast to orientable bonds) having the fixed(!)
direction from, say, $n_i$ to $n_k$ and vice versa.
b) Considering the whole context rather from the viewpoint of
(discrete) manifolds, we defined the $\di$ as linear forms, mapping
the {\it tangential basis vectors} $\partial_{ik'}$ attached to the node
$n_i$ onto $\delta_{kk'}$, i.e:
\beq <d_{ki}|\partial_{ik'}>=\delta_{kk'}\eeq
In this sense the $\di,\dk$ can be regarded as objects being attached
to the nodes $n_i,n_k$ respectively, while the bonds $\bi$ are
delocalized, living in the ''environment'' between the nodes. We think
that this property is of physical significance and the crucial
difference between $\di$ and $\bi$.

In this new basis the {\it differential} $df$ of Observation 2.4
acquires the form\\[0.5cm]
{\bf 2.6 Corollary}: \beq df=\sum_{ik}(f_k-f_i)\di\eeq
{\bf 2.7 Lemma}: It is obvious that the map $\delta$ can be
canonically extended to this larger space by:
\beq \delta_1:\;\di\to\;n_k\eeq
A last but important point we want to mention is Observation 3.13 of
\cite{1}:\\[0.5cm]
{\bf 2.8 Observation(Graph-Laplacian)}:
\beq
\delta df=-\sum_i(\sum_kf_k-v_i\cdot f_i)n_i=-\sum_i(\sum_k(f_k-f_i))n_i
=:-\Delta f\eeq
where $v_i$ denotes the {\it node(vertex) degree} or the {\it valency}
of the node $n_i$, i.e. the number of nearest neighbors $n_k$ being
connected to it by a bond.

It is interesting that this graph laplacian, which we developed
following a completely different line of reasoning in \cite{1}, is
intimately connected with an object wellknown to graph theorists,
i.e. the {\it adjacency matrix} of a graph.\\[0.5cm]
{\bf 2.9 Definition(Adjacency Matrix)}: i)The entries $a_{ik}$ of the
{\it adjacency matrix} $A$ have the value one if the nodes $n_i,n_k$
are connected by a bond, zero elsewhere. If the graph is {\it
  undirected} (but orientable; the case we mainly discuss), the
relation between $n_i,n_k$ is {\it symmetric}, i.e.
\beq a_{ik}=1\quad\Rightarrow\quad a_{ki}=1\quad\mbox{etc.}\eeq
with the consequence:\\
ii) If the graph is {\it simple} and {\it undirected}, $A$ is a
symmetric matrix with zero diagonal elements.\\[0.5cm]
Remark: More general $A$'s occur if more general graphs are
discussed.\\[0.5cm]
{\bf Observation}: With our definition of $\Delta$ it holds:
\beq \Delta=A-V\eeq
where $V$ is the diagonal {\it degree matrix}, having $v_i$ as
diagonal entries.\\[0.5cm]
Proof: As we have not yet introduced a Hilbert space structure (which
we will do below), the proof has to be understood, for the time being,
in an algebraic way. We then have:
\beq Af=A(\sum
f_in_i)=\sum_i(\sum_{k-i}n_k)=\sum_i(\sum_{k-i}f_k)n_i\eeq
\beq Vf=\sum_i(v_if_i)n_i\eeq
hence the result.\\[0.5cm]
Remarks:i) Here and in the following we use the abbreviation $k-i$ if
the nodes $n_k,n_i$ are connected by a bond, the summation always
extending over the first variable.\\
ii) From this interplay between {\it graph geometry} and {\it
  functional analysis} follow a lot of deep and fascinating results,
as is always the case in mathematics if two seemingly well separated
fields turn out to be closely linked on a deeper level. This is
particularly the case if geometry is linked with algebra or functional
analysis.
\section{Some Functional Analysis on Graphs}
After the preliminary remarks made in the previous section we now
enter the heart of the matter. Our first task consists of endowing a
general graph with both a sufficiently reach and natural Hilbert space
structure on which the various operators to be constructed in the
following can act.\\[0.5cm]
{\bf 3.1 Definition (Hilbert Space)}: i)In $C_0$ we choose the subspace
$\it H_0$ of sequences $f$ so that:
\beq \|f\|^2=\sum |f_i|^2<\infty\eeq
ii) In $C_1,C_1'$ respectively we make the analogous choice:
\beq {\it H_1}:=\{g|\,\|g\|^2:=\sum
|g_{ik}|^2<\infty;\,g_{ik}=-g_{ki}\}\eeq
\beq {\it H_1'}:=\{g'|\,\|g'\|^2:=\sum |g'_{ik}|^2<\infty\}\eeq
with $g=\sum g_{ik}d_{ik},\;g'=\sum g'_{ik}d_{ik}$ and the respective
ON-bases $\{n_i\},\,\{d_{ik}\}$, that is
$<d_{ik}|d_{i'k'}>=\delta_{ii'}\delta_{kk'}$.\\[0.5cm]
Remark: The convention in ii) is made for convenience in order to
comply with our assumption $b_{ik}=-b_{ki}$, which is to reflect that
the bonds $b$ are undirected and that functions over it should be
given modulo their possible orientation! Members of ${\it H_1}$ can
hence also be written
\beq \sum g_{ik}d_{ik}=1/2\sum g_{ik}d_{ik}+1/2\sum
g_{ki}d_{ki}=1/2\sum g_{ik}(d_{ik}-d_{ki})=1/2\sum g_{ik}b_{ik}\eeq
iii) As $\it H,{\it H'}$ we take the {\it direct sums}:
\beq {\it H}:={\it H_0}\oplus{\it H_1},\;{\it H'}:={\it H_0}\oplus{\it
  H'_1}\eeq
{\bf 3.2 Observation}: Obviously $\Hib$ is a subspace of $\Hsb$ and we
have
\beq <b_{ik}|b_{ik}>=2\eeq
i.e. the $b_{ik}$ are not(!) normalized if the $d_{ik}$ are. We could
of course enforce this but then a factor two would enter
elsewhere.\vspace{0.5cm}

With these definitions it is now possible to regard the maps
$d,\,\delta$ as full-fledged operators between these Hilbert
(sub)spaces.\\[0.5cm]
{\bf 3.3 Assumption}: To avoid domain problems and as it is natural
anyhow, we assume from now on that the {\it node degree} $v(n_i)$ is
{\it uniformly bounded} on the graph $G$, i.e.
\beq v_i<v_{max}\;\mbox{for all}\;i\eeq
{\bf 3.4 Definition/Observation}: i)
\beq d:\;\Hia\to\Hib,\quad \delta:\;\Hib\to\Hia\eeq
ii) $d_{1,2}$ with
\beq d_{1,2}:\;n_i\to\sum d_{ki},\;\sum d_{ik}\eeq
respectively and linearly extended, are operators from $\Hia\to\Hsb$
and we have
\beq d=d_1-d_2\eeq
iii) $\delta$ may be extended in a similar way to $\Hsb$ via:
\beq \delta_{1,2}:\;d_{ik}\to n_k,\,n_i\eeq
respectively and linearly extended. We then have:
\beq
\delta_1(b_{ik})=\delta(b_{ik})=n_k-n_i=(\delta_1-\delta_2)(d_{ik})\eeq

It is remarkable that $v_i\le v_{max}$ implies that all the above
operators are {\it bounded}(!) (in contrast to similar operators in
the continuum, which are typically unbounded).\\[0.5cm]
{\bf 3.5 Theorem}: All the operators introduced above are bounded on
the respective Hilbert spaces, i.e. their domains are the full Hilbert
spaces under discussion.\\[0.5cm]
Proof: We prove this for, say, $d$; the other proofs are more or less
equivalent.
\beq d:\,\Hia\ni\sum
f_in_i\to\sum_i(\sum_{k=1}^{v_i}b_{ki})=\sum_{ik}(f_k-f_i)d_{ik}\eeq
and for the norm of the rhs:
\beq
\|rhs\|^2=\sum_{ik}|(f_k-f_i)|^2=\sum_{ik}(|f_i|^2+|f_k|^2-\overline{f_k}f_i-\overline{f_i}f_k)=2\cdot\sum_i
v_i|f_i|^2-2\cdot\sum_{ik}\overline{f_k}f_i\eeq
The last expression can be written as:
\beq \|df\|^2=2(<f|Vf>-<f|Af>)=<f|-2\Delta f>\eeq
which is a remarkable result. It shows, among other things, that $d$
and its norm are closely connected with the {\it expectation values}
of the {\it adjacency} and {\it degree matrix} respectively the {\it
  graph Laplacian} (introduced in the previous section). This is, of
course, no accident and relations like these will be clarified more
systematically immediately.

It follows already from the above that we have:\\[0.5cm]
{\bf 3.6 Observation}: 
\beq \|df\|^2=<f|d^{\ast}df>=<f|-2\Delta f>\eeq
i.e.
\beq
d^{\ast}d\,=\,-2\Delta\;\mbox{and}\;\|d\|^2=\sup_{\|f\|=1}<f|-2\Delta
f>=\|-2\Delta\|\eeq
We then have:
\beq 0<\sup_{\|f\|=1}<f|-2\Delta f>\leq
2v_{max}+2\sup_{\|f\|=1}|<f|Af>|\eeq

Due to our assumption $A$ is a (in general infinite) hermitean matrix
with entries $0$ or $1$, but with atmost $v_{max}$ nonzero entries in
each row and vanishing diagonal elements.\\[0.5cm]
Remark: i)It is possible to treat also more general $A$'s if we admit
more general graphs (e.g. socalled {\it multi graphs}; see \cite{8}
and \cite{9}).\\
ii) It should be noted that, whereas $A$ is a matrix with nonnegative
entries, it is not(!) {\it positive} in the sense of {\it linear
  operators}. The positive(!) operator in our context is the Laplacian
$-\Delta$, which can be seen from the above representation as
$1/2d^{\ast}d$. On the other side matrices like $A$ are frequently
called positive in the matrix literature, which is, in our view,
rather missleading.\\
iii) It should further be noted that we exclusively use the {\it
  operator norm} for matrices (in contrast to most of the matrix
literature), which may also be called the {\it spectral norm}. It is
unique insofar as it coincides with the socalled {\it spectral radius}
(cf. e.g. \cite{11} or \cite{12}), that is
\beq \|A\|:=\sup\{|\lambda|;\,\lambda\in spectr(A)\}\eeq

After these preliminary remarks we will now estimate the norm of
$A$. To this end we exploit a simple but effective inequality, which
perhaps better known in numerical mathematics and which we adapt to
infinite matrices of the above type.\\[0.5cm]
{\bf 3.7 Theorem (Variant of Gerschgorin Inequality)}: Let $A$ be a
finite adjacency matrix with admost $v_{max}$ nonzero entries in each
row. Then:
\beq \sup\{|\lambda|;\,\lambda\;\mbox{an eigenvalue}\}\leq v_{max}\eeq
Proof: This is an immediate application of the original Gerschgorin
inequality to our case (vanishing diagonal elements; see
e.g. \cite{13}).\vspace{0.5cm}

To treat the infinite case with finite $v_{max}$, we choose a sequence
of n-dimensional subspaces $X_n$ with basis elements $e_1,\ldots
e_n$. In these subspaces the corresponding projections $A_n$ of the
infinite $A$ fulfills the above assumption. We then have for a
normalizable vector $x;\,\sum |x_i|^2<\infty$ that
\beq <x|A_nx>\to<x|Ax>\eeq
The same does then hold for functions of $A,\,A_n$, in particular for
the spectral projections; hence:\\[0.5cm]
{\bf 3.8 Theorem (Norm of $A$)}: With the adjacency matrix $A$
possibly infinite and a finite $v_{max}$ we have the important
estimate:
\beq \|A\|=\sup\{|\lambda|;\,\lambda\in\;spectr(A)\}\leq v_{max}\eeq
This result concludes also the proof of theorem 3.5!\vspace{0.5cm}

The above reasoning shows that many of the technical difficulties
(e.g. domain problems) are absent on graphs with uniformly bounded
degree as all the operators turn out to be bounded under this premise.
We want to conclude this section with deriving some relations among
the operators introduced in Definition/Observation 3.4:

We already realized that
\beq d^{\ast}d=-2\Lambda\eeq
holds, but we did not make explicit the true nature of
$d^{\ast}$. On the one side
\beq d:\,\Hia\to\Hib\;,\;d^{\ast}:\,\Hib\to\Hia\eeq
On the other side 
\beq \delta:\,\Hib\to\Hia\eeq
and for $g\in\Hib$, i.e. $g_{ik}=-g_{ki}$:
\beq
\sum_{ik}g_{ik}d_{ik}=1/2\sum_{ik}g_{ik}(d_{ik}-d_{ki})=1/2\sum_{ik}g_{ik}b_{ik}\eeq
Calculating now
\beq <g|df>=\sum g_{ik}(f_k-f_i)=<2\delta g|f>\eeq
we can infer:\\[0.5cm]
{\bf 3.9 Observation}: i)The adjoint $d^{\ast}$ of $d$ with respect to
the spaces $\Hia,\Hib$ is $2\delta$, which also follows from the
comparison of the representation of the graph Laplacian in Observation
2.8 and Definition/Observation 3.4, i.e:
\beq \delta d=-\Delta=1/2d^{\ast}d\eeq
ii) For the natural extensions $d_{1,2},\delta_{1,2}$ of $d,\delta$ to the larger (sub)spaces
$\Hia,\Hsb$ we have (cf. the definitions in Definition/Observation
3.4):
\beq \delta_1=(d_1)^{\ast}\;,\;\delta_2=(d_2)^{\ast}\eeq
hence 
\beq (\delta_1-\delta_2)=(d_1-d_2)^{\ast}=d^{\ast}\eeq
Proof of ii):
\beq <g|d_1f>=\sum_{ik}g_{ik}f_i=<\delta_1g|f>\quad\mbox{etc.}\eeq

In other words, the important result for the following is that with
respect to the larger space $\Hsb$ the adjoint $d^{\ast}$ of $d$ is
$(\delta_1-\delta_2)$ and we have:
\beq (\delta_1-\delta_2)b_{ik}=2(n_k-n_i)=2\delta b_{ik}\eeq
Henceforth we identify $d^{\ast}$ with $(\delta_1-\delta_2)$.

\section{The Spectral Triple on a general Graph}
The Hilbert space under discussion in the following is
\beq \Hi=\Hia\oplus\Hsb\eeq
The {\it natural representation} of the function algebra ${\cal F}$
\beq \{f;f\in {\cal C}_0,\sup_i |f_i|<\infty\}\eeq
on $\Hi$ is given by:
\beq \Hia:\;f\cdot f'=\sum f_if'_i\cdot n_i\;\mbox{for}\;f'\in \Hia
\eeq
\beq \Hsb:\;f\cdot\sum g_{ik}d_{ik}:=\sum f_ig_{ik}d_{ik}\eeq
From previous work (\cite{1}) we know that ${\cal C}'_1$ carries also
a right-module structure, given by:
\beq \sum g_{ik}d_{ik}\cdot f:=\sum g_{ik}f_k\cdot d_{ik}\eeq
Remark: For convenience we do not distinguish notationally between
elements of $\Fc$ and their Hilbert space
representations.\vspace{0.5cm}

The important and nontrivial object is the socalled {\it Dirac
  operator} $D$. As $D$ we will take the operator:
\beq D:=\left( \begin{array}{cc}0 & d^{\ast}\\d & 0 \end{array} \right)\eeq
acting on
\beq \Hi=\left( \begin{array}{c}\Hia \\ \Hsb \end{array} \right)\eeq
with
\beq d^{\ast}=(\delta_1-\delta_2)\eeq 
{\bf 4.1 Definition/Observation (Spectral Triple)}: Our {\it spectral
  triple} on a general graph is given by
\beq (\Hi,\Fc,D)\eeq
introduced in the preceeding formulas.

As can be seen from the above, the connection with the graph Laplacian
is relatively close since:
\beq D^2=\left( \begin{array}{cc}\dast d & 0\\0 & d\dast \end{array}
\right)\eeq
and
\beq \dast d=-2\Delta\eeq
$d\dast$ is the corresponding object on $\Hsb$.\vspace{0.5cm}

We now calculate the commutator $[D,f]$ on an element $f'\in\Hia$:
\beq (d\cdot f)f'=\sum_{ik}(f_kf'_k-f_if'_i)d_{ik}\eeq
\beq (f\cdot d)f'=\sum_{ik}f_i(f'_k-f'_i)d_{ik}\eeq
hence
\beq [D,f]f'=\sum_{ik}(f_kf'_k-f_if'_k)d_{ik}\eeq
On the other side the right-module structure yields:
\beq df\cdot f'=(\sum_{ik}(f_k-f_i)d_{ik})\cdot (\sum_k f'_kn_k)=\sum_{ik}(f_kf'_k-f_if'_k)d_{ik}\eeq
In a next step one has to define $df$ as operator on $\Hsb$. This is
done in the following way (the reason follows below):
\beq df:\;d_{ik}\to (f_i-f_k)n_k\eeq
and linear extension.\\[0.5cm]
{\bf 4.2 Definition}: The representation of $df$ on $\Hi$ is defined
in the following way:
\beq df:\;\Hia\to\Hsb\;,\;\Hsb\to\Hia\quad\mbox{via}\eeq
\beq n_k\to(f_k-f_i)d_{ik}\;,\;d_{ik}\to (f_i-f_k)n_k\eeq
We are now able to calculate the commutator $[D,f]$ on $\Hsb$:
\beq (\dast\cdot f)g-(f\cdot\dast)g=\sum (f_i-f_k)g_{ik}\cdot n_k\eeq
with $g=\sum g_{ik}d_{ik}\in\Hsb$.\\
On the other side:
\beq (df)g=\sum g_{ik}(df)d_{ik}=\sum (f_i-f_k)g_{ik}\cdot n_k\eeq
We hence have the important result:\\[0.5cm]
{\bf 4.3 Theorem (Dirac Operator)}: With the definition of the
representation of $df$ on $\Hi$ as in Definition 4.2 it holds:
\beq [D,f]x=(df)x\;,\;x\quad\mbox{being an element of}\quad\Hi\eeq
and in particular:
\beq [d,f]=df|_{\Hia}\;,\;[\dast,f]=df|_{\Hsb}\eeq
with the two maps intertwining $\Hia$ and $\Hsb$. \\[0.5cm]
Remark: In the above sections we have discussed a Hilbert space representation
based on the Hilbert space $\Hia\oplus \Hsb$ and operators $d,\dast$
respectively $df$ etc. In the same way one could choose a '{\it dual
  representation}' over the '{\it tangential space}' built from
the $\partial_{ik}$'s (introduced in \cite{1}; see also formula (6) in
Definition/Observation 2.5). In this case the operator $d$ goes over
into the dual object $\nabla$. This choice is natural as well and
arises from a '{\it dualization}' of the above one. It will be
discussed elsewhere in connection with the developement of a discrete
Euler-Lagrange formalism on graphs.

\section{The Connes-Distance Function on Graphs}
The first step consists of calculating the norm of $[D,f]=df$
respectively its supremum under the condition $\|f\|\leq 1$. We have
(with the help of previous calculations):
\beq [D,f]=\left( \begin{array}{cc} 0   & [\dast,f] \\ 
                                  {[}d , f] &  0          \end{array}
                              \right) 
\eeq
and with $\{f_i\}$ uniformly bounded and real, i.e. $f$ a bounded and
s.a. operator:
\beq [d,f]^{\ast}=-[\dast,f]\eeq
From the general theory we have:
\beq \|A\|=\|A^{\ast}\|\eeq
(where here and in the following $A$ denotes a general operator and
not(!) the adjacency matrix) hence\\[0.5cm]
{\bf 5.1 Observation}:
\beq \|[d,f]\|=\|[\dast,f]\|\eeq
and
\beq \|[D,f](X)\|^2=\|[d,f]x\|^2+\|[\dast,f]y\|^2\eeq
with
\beq X:=\left( \begin{array}{c}x \\ y \end{array} \right)\eeq 

Choosing the abbreviation $A:=[d,f]$, the norm of $[D,f]$ is:
\beq
\|[D,f]\|^2=\sup\{\|Ax\|^2+\|A^{\ast}y\|^2;\,\|x\|^2+\|y\|^2=1\}\eeq
Normalizing $x,y$ to $\|x\|=\|y\|=1$ and representing a general
normalized vector $X$ as:
\beq X=\lambda x+\mu
y\;,\;\lambda,\mu>0\;\mbox{and}\;\lambda^2+\mu^2=1 \eeq
we get:
\beq
\|[D,f]\|^2=\sup\{\lambda^2\|Ax\|^2+\mu^2\|A^{\ast}y\|^2;\|x\|=\|y\|=1,\lambda^2+\mu^2=1\}\eeq
where now $x,y$ can be varied independently of $\lambda,\mu$ in
their respective admissible sets, hence:\\[0.5cm]
{\bf 5.2 Conclusion}: 
\beq \|[D,f]\|^2=\|A\|^2=\|[d,f]\|^2=\|[\dast,f]\|^2\eeq
where the operators on the rhs act in the reduced spaces $\Hia,\Hsb$
respectively.\vspace{0.5cm}

It follows that in calculating $\|[D,f]\|$ one can restrict oneself to
the easier to handle $\|[d,f]\|$. For the latter expression we then get
from the above ($x\in\Hia$):
\beq \|df\cdot x\|^2=\sum_i(\sum_{k=1}^{v_i}(f_i-f_k)^2)|x_i|^2\eeq
Abbreviating 
\beq \sum_{k=1}^{v_i}(f_k-f_i)^2=:a_i\geq 0\eeq
and calling the supremum over $i$ $a_s$, it follows:
\beq \|df\cdot x\|^2= a_s\cdot(\sum_i a_i/a_s\cdot|x_i|^2)\leq
a_s\eeq
for $\|x\|^2=\sum_i |x_i|^2=1$.

On the other side, choosing a sequence of normalized basis vectors $x_{\nu}$ so
that the corresponding $a_{\nu}$ converges to $a_s$ we get:
\beq \|df\cdot x_{\nu}\|^2\to a_s\eeq
{\bf 5.3 Theorem (Norm of $\|[D,f]\|$)}:
\beq \|[D,f]\|=\sup_i(\sum_{k=1}^{v_i}(f_k-f_i)^2)^{1/2}\eeq

The '{\it Connes-distance function}' is now defined as
follows:\\[0.5cm]
{\bf 5.4 Definition (Connes-distance function)}:
\beq dist_C (n,n'):=\sup\{|f_{n'}-f_n|;\|[D,f]\|=\|df\|\leq 1\}\eeq
where $n,n'$ are two arbitrary nodes on the graph.\\[0.5cm]
Remark: It is easy to prove that this defines a metric on the
graph.\\[0.5cm]
{\bf 5.5 Corollary}: It is sufficient to vary only over the set
$\{f;\|f\|=1\}$.\\[0.5cm]
Proof: This follows from
\beq |f_k-f_i|=c\cdot|f_k/c-f_i/c|\;;\;c=\|df\|\eeq
and
\beq \|d(f/c)\|=c^{-1}\|df\|=1\eeq
with $c\leq1$ in our case.\vspace{0.5cm}

It is evidently a nontrivial task to calculate this distance on an
arbitrary graph as the above constraint is quite subtle whereas it is
given in a closed form. We refrain at this place from a complete
discussion but add only the following remarks concerning the connection to the ordinary distance function
introduced in the beginning of the paper.

Having an admissible function $f$ so that
$\sup_i(\sum_{k=1}^{v_i}(f_k-f_i)^2)^{1/2}\leq 1$, this implies that,
taking a '{\it minimal path}' $\gamma$ from, say, $n$ to $n'$, the
jumps $|f_{\nu+1}-f_{\nu}|$ between neighboring nodes along the path
have to fulfill:
\beq |f_{\nu+1}-f_{\nu}|\leq 1\eeq
and are typically strictly smaller than $1$.

On the other side the Connes distance would only become identical to the
ordinary distance $d(n,n')$ if there exist a sequence of admissible
node functions with these jumps approaching the value $1$ along such a
path. Only in this case one would get:
\beq \sum_{\gamma}|f_{\nu+1}-f_{\nu}|\to
\sum_{\gamma}1=length(\gamma)\eeq
The construction of such functions is however an intricate and
''nonlocal'' business if at the same time the above constraint
concerning the jumps between neighboring nodes is to be fulfilled.\\[0.5cm]
{\bf 5.6 Observation (Connes-distance)}: In general one has the
inequality
\beq dist_C(n,n')\leq d(n,n')\eeq

This result can e.g. be tested in a simple example. Take, say, a
square with vertices and edges:
\beq x_1-x_2-x_3-x_4-x_1\eeq
Let us calculate the Connes-distance between $x_1$ and $x_3$. 

As the $\sup$ is taken over functions(!) the summation over elementary
jumps is (or rather: has to be) pathindependent (this is in fact both a subtle and crucial
constraint for practical calculations). It is an easy exercise to see
that the $sup$ can be found in the class where the two paths between
$x_1,x_3$ have the '{\it valuations}' ($1\geq a\geq0$):
\beq x_1-x_2:\,a\;,\;x_2-x_3:\,(1-a^2)^{1/2}\eeq
\beq x_1-x_4:\,(1-a^2)^{1/2}\;,\;x_4-x_3:\,a\eeq
Hence one has to find $\sup_{0\leq a\leq1}(a+\sqrt{1-a^2})$. Setting
the derivative with respect to $a$ to zero one gets
$a=\sqrt{1/2}$. Hence:\\[0.5cm]
{\bf 5.7 Example (Connes-distance on a square)}:
\beq dist_C(x_1,x_3)=\sqrt{2}<2=d(x_1,x_3)\eeq

\end{document}